# The evolution of Materials Acceleration Platforms - towards the laboratory of the future with AMANDA

Jerrit Wagner[1,2,*], Christian G. Berger[1,2], Xiaoyan Du[1,2], Tobias Stubhan[3], Jens A. Hauch[1,2,*], Christoph J. Brabec[1,2]

[1] *Forschungszentrum Jülich GmbH, Helmholtz-Institute Erlangen-Nürnberg for Renewable Energy (IEK-11), Immerwahrstraße 2, 91058 Erlangen, Germany*
[2] *Institute of Materials for Electronics and Energy Technology (i-MEET), Department of Materials Science and Engineering, Friedrich-Alexander University Erlangen-Nürnberg, Martensstrasse 7, 91058 Erlangen, Germany*
[3] *SCIPRIOS GmbH, Fürther Str. 244a, 90429 Nürnberg, Germany*
[*] *Authors to whom correspondence should be addressed: j.wagner@fz-juelich.de and j.hauch@fz-juelich.de*

## Abstract

The development of complex functional materials poses a multi-objective optimization problem in a large multidimensional parameter space. Solving it requires reproducible, user independent laboratory work and intelligent preselection of experiments. However, experimental materials science is a field where manual routines are still predominant, although other domains like pharmacy or chemistry have long used robotics and automation. As the number of publications on Materials Acceleration Platforms (MAPs) increases steadily, we review selected systems and fit them into the stages of a general material development process to examine the evolution of MAPs. Subsequently we present our approach to laboratory automation in materials science. We introduce *AMANDA* (***A****utonomous **M**aterials **an**d **D**evice **A**pplication Platform*)[1], a generic platform for distributed materials research comprising a self-developed software backbone and several MAPs. One of them, *LineOne (L1)*, is specifically designed to produce and characterize solution processed thin-film devices like organic solar cells (OSC). It is designed to perform precise closed-loop screenings of up to 272 device variations per day yet allows further upscaling. Each individual solar cell is fully characterized and all process steps are comprehensively documented. We want to demonstrate the capabilities of *AMANDA L1* with OSCs based on PM6:Y6 with 13.7% efficiency when processed in air. Further we discuss challenges and opportunities of highly automated research platforms and elaborate on the future integration of additional techniques, methods and algorithms in order to advance to fully autonomous self-optimizing systems - a paradigm shift in functional materials development leading to the laboratory of the future.

---

[1] www.amanda-platform.com

# Introduction

Industrial production is based on a continual repetition of the same production process. Since the 18th century, increasingly automated machines have been developed to speed up production and improve quality. Today, the ever-growing automation of industrial processes has led to completely automated production plants. Computational methods and data collection further integrate the knowledge about the manufacturing processes into production planning systems in order to make informed decisions [1]. As such, these "smart" industrial processes are referred to as "Industry 4.0" [2].

The automation of simple scientific tasks started much later in the late 19$^{th}$ century with simple tasks like washing filtrates or conducting solvent extraction [3]. In the laboratory environment it was not until the 1980s with the pioneering work of Sasaki that more complex setups were implemented, where several devices were coupled together and samples were automatically handled [4, 5]. Until today, these advances are mainly found in the field of clinical and pharmacological research [3, 6].

The potential for automation as a disruptive technology to greatly accelerate materials science has been recognized for years [7]. Several programs like the Materials Genome Initiative [8] (MGI) and the Clean Energy Materials Innovation Challenge [9] target faster material development and commercialization by introducing automation into the field of materials science. Latter defined the combination of highly automated robotic setups with inline characterization and artificial intelligence (AI) as a Materials Acceleration Platform (MAP). First approaches of automation have been shown already a decade ago, mainly utilizing high-throughput techniques [10–12]. AI guided selection of experimental parameters, so-called closed-loop optimization, emerged only recently [13–15].

Coupling modular robotic techniques with comprehensive data collection and artificial intelligence enables new experimental approaches like the inverse design of materials synthesis [16]. Automation can also be the gateway to higher reproducibility, an attribute often missing in scientific research. According to a survey by Baker among 1500 physics and chemistry scientists, more than 40% have failed to reproduce their own experiments. This rate is even as high as 60% when trying to reproduce the experiments from other laboratories [17]. The use of automated devices such as robots or pipetting units reduces the error into defined margins, because of enhanced repeatability and their constant accuracy. The specified operational ranges can be taken into consideration during process design but also evaluation. Automated research platforms can therefore contribute to increase the reproducibility of, and thus the confidence in published data.

While classical automation in mass production is based on the strict repetition of the same process, it is characteristic for laboratory research in materials science to vary parameters and the process itself in order to understand and optimize a material system. For the automation of a laboratory it is therefore critical to keep the workflows as flexible as possible. However, automated, robot-supported processes always lead to restrictions. While a typical robot for handling samples has between 3 and 10 joints, the human body has several hundred joints [18, 19]. This means that the automated system will have significantly fewer degrees of freedom than a researcher operating with the same processing and characterization equipment. In addition humans are not confined to a particular area - they can walk to different parts of the laboratory, while the typical robot is fixed in location. First approaches are being shown, like the *mobile robot chemist,* to compensate for this weakness [13]. Nevertheless a researcher can also make ad-hoc decisions to change a workflow or introduce a new procedure at any time, while in an automated system the development and testing of new features often delays utilization.

Despite these challenges flexible automated setups with customizable processes have the potential to encourage the transition from traditional lab work into the laboratory of the future. Highly customizable experimental design preserves a researcher's ability of individually adapting the experiments and is key for scientific discoveries. Our vision for a lab of the future expands laboratory work with automation for enhanced experimental precision, implements digital technologies for data collection and storage, performs simulations and uses artificial intelligence to guide the researcher.

In this publication we examine the evolution of Materials Acceleration Platforms and introduce a system that goes significantly beyond the state-of-the-art. Our *Autonomous Materials and Device Application Platform: AMANDA*[2] is designed as a framework for the laboratory of the future, including a generic software backbone with the capability of controlling multiple Materials Accelerations Platforms. The software allows for a high degree of automation to accelerate research while keeping the lab processes flexible. Data is collected at all stages of the process; during preparation, execution and characterization for a comprehensive and complete documentation. All data sets are interlinked and retrievable to allow systematic analyses across experiments. Self-driven feedback loops additionally enable autonomous experimentation against optimization targets. Additionally, the system implements an experiment-as-a-service (EaaS) approach. Researchers define their experiments through a web-interface. The processing is then conducted automatically on *AMANDA*'s Materials Acceleration Platforms and the results can again be retrieved online.

*AMANDA* is used in this publication to drive an automated experimental line which is designed to accelerate the materials and process development for solution processed organic solar cells - *LineOne (L1)*. Organic photovoltaics (OPV) is a promising field of materials science and a viable path towards cleaner energy production [20]. Research is driven by new materials and the production routes are defined by the complex bulk heterojunction morphology, leading to a large variety of process parameters influencing the solar cell performance [21–27]. Lately the PM6[3]:Y6[4] material system attracted a lot of attention in the OPV research community for boosting the efficiency of organic solar cells up to 18% [25, 28–31]. Adding further materials, like additional electron donors or acceptors into the PM6:Y6 active layer mixture can improve both efficiency [32, 33] and long-term stability [34] of produced solar cell devices. However, the complexity in the fabrication process as well as the grown number of process parameters in multi-component systems makes it increasingly difficult to investigate such systems manually [32, 35]. For a systematic screening of the vast amount of promising OPV materials and combinations, utilizing automation is a viable solution and becomes more and more necessary to reduce the time of material qualification, as the availability of new materials is increasing steadily [36].

---

[2] www.amanda-platform.com

We demonstrate the capabilities of *L1* at the hand of the state-of-the-art OPV material system PM6:Y6 [23, 25]. In this work *L1* processed complete PM6:Y6 based bulk heterojunction solar cells in air. A power conversion efficiency of 13.7% was reached, which compares to other reports on the processing of that material system in ambient atmosphere [37]. Furthermore, a reproducibility study was performed over the course of three months on that material system. In 19 different experiments with the same materials and process parameters we show that *L1* performs steadily at a very low deviation with an interquartile range of 0.74% in power conversion efficiency (*PCE*). We take advantage of this reproducibility to investigate the influence of solid content and solution amount for the spin coating process on the solar cell performance. Additionally we discuss the difficulties but also the advantages of operating fully integrated MAPs. Finally, we elaborate the future development roadmap for the *AMANDA* Platform as well as the opportunities we see by the implementation of such systems.

# The evolution of Materials Acceleration Platforms

The development of humankind is strongly linked to the evolution of materials. As knowledge and technical skills improved, people gained access to more complex material compositions and manufacturing processes. Figure 1a qualitatively relates the complexity of material composition and processes for fabricating functional materials throughout the historical eras. Material compositions used in the distant past, such as Bronze, consisted of only two elements. In strong contrast to this binary compound, the chemical parameter space

---

[3] (poly[(2,6-(4,8-bis(5-(2-ethylhexyl)-4-fluorothiophen2-yl)benzo[1,2-b:4,5-b0]dithiophene))-co-(1,3-di(5-thiophene-2-yl)-5,7- bis(2-ethylhexyl)-benzo[1,2-c:4,5-c0 ]dithiophene-4,8-dione))] - chemical structure is given in figure 5a

[4] 2,2′-((2Z,2′Z)-((12,13-bis(2-ethylhexyl)-3,9-diundecyl-12,13-dihydro-[1,2,5]thiadiazolo[3,4-e]thieno[2,"3′′:4',5′]thieno[2′,3′:4,5]pyrrolo[3,2-g]thieno[2′,3′:4,5]thieno[3,2-b]indole-2,10-diyl)bis(methanylylidene))bis(5,6-difluoro-3-oxo-2,3-dihydro-1H-indene-2,1-diylidene))dimalononitrile) - chemical structure is given in figure 5a

available in the organic chemistry of small molecules alone is estimated at around $10^{63}$ possible molecules [38]. A similar increase of complexity is seen in the processing of materials. To make bronze, tin and copper ore were molten at around 1000°C for a few hours about 5000 years ago [39]. Nowadays, the process to get electronic grade, high purity silicon is much more complex. The commonly used Czochralski process for the production of silicon single crystals, needs temperatures of around 1400°C, high purity raw material containing e.g. less than 0.2 parts per billion of boron and phosphorus, intensive Argon purging of the reactor and slow, yet very consistent pulling speeds of a single crystal seed of few millimeter per minute [40]. The dimensionality of the parameter space increases dramatically with growing complexity of the compounds and processes. A vast multi-dimensional space requires a structured examination, however finding its optimum becomes an increasingly daunting task. In order to master this challenge, materials science today relies on the intuition and experience of scientists to reduce the necessary number of samples. In order to transfer this complex optimization problem to automated systems, it is crucial to understand the process of material development and divide it into its individual parts. The commonly practiced materials discovery procedure to develop and optimize new functional materials can be reduced to the following steps:

1. Formulating a hypothesis or research question
2. Selection of "precursor" materials
3. Performing a process with the selected precursors to generate a product
4. Characterizing the properties of the product by measurement(s)
5. Evaluating the overall result based on the characterization

Conventional laboratory research in materials science experimentation is predominantly manual. The individual experimental parameters are chosen by the researcher (oftentimes students) and varied in order to reach some sort of optimum. The selection of parameters relies on a mixture of knowledge, experimentation results and intuition in order to improve performance step by step (compare figure 1c: manual research). Precision relies on the thoroughness, diligence, dexterity and aptitude of the scientist performing the work, all factors that may change on a daily basis. This leads to a strong user-dependency of experimental results and can therefore impair the reproducibility of scientific results. Furthermore, only successful experiments and results are usually published while unsuccessful experiments are often not even documented. Much of the knowledge that might still be important for a systematic understanding is often lost as it has neither been stored nor shared among peers.

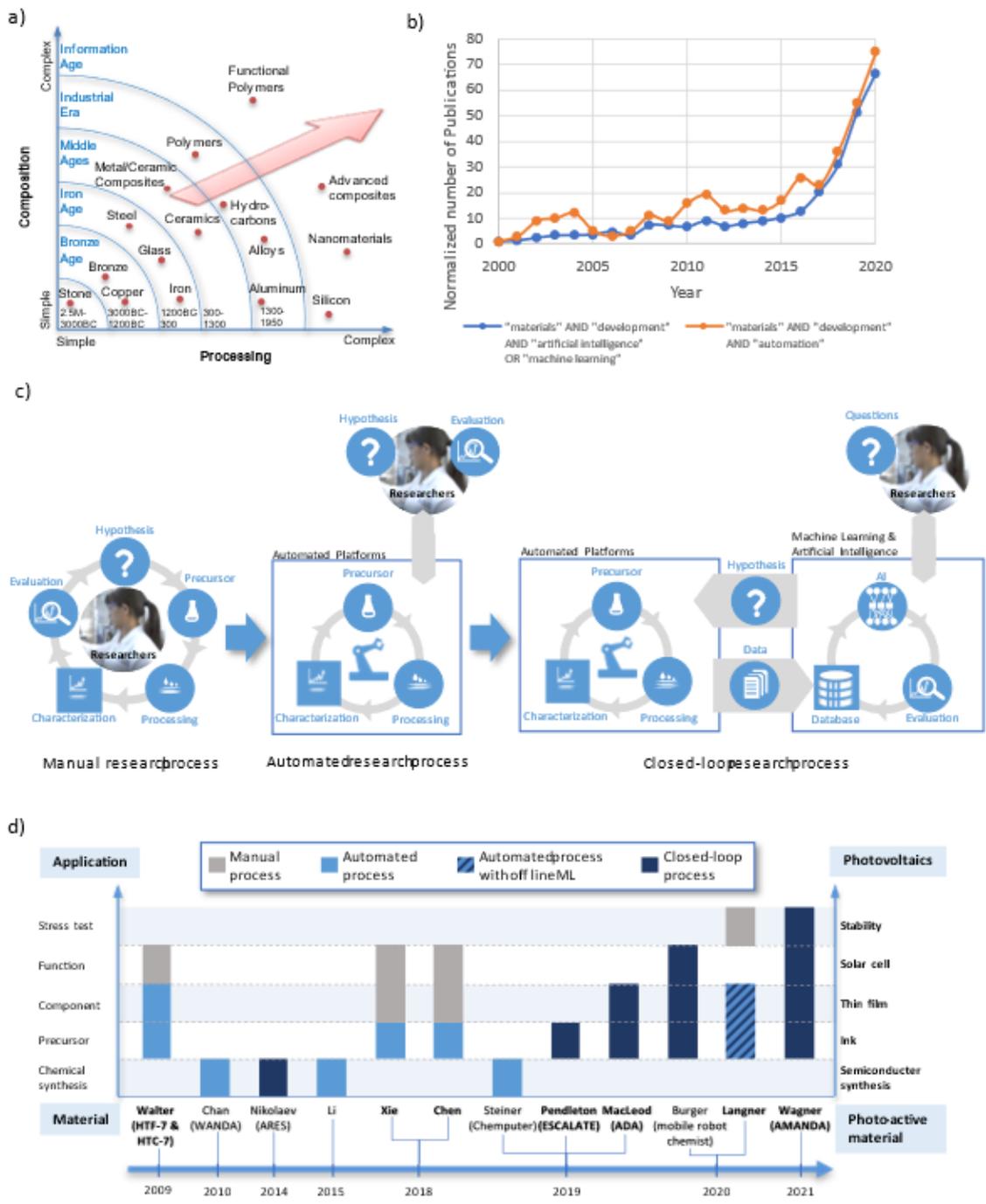

*Fig. 1: a) Qualitative representation of processing and composition complexity of various functional materials in a historical perspective. b) Time development of the number of publications found with the search terms "materials" AND "development" AND ("artificial intelligence" OR "machine learning"), and "materials" AND "development" AND "automation" in publication title, abstract and keywords. The numbers of publications per year are gathered from Scopus and are normalized to the year 2000. c) Schematic view of materials discovery process in the classical approach, automated and in AI enhanced closed-loop mode. d) Comparison of different MAPs of the last decade, with MAPs focusing on photovoltaic materials in bold. The y-axis shows the capability of the systems categorized*

*in different stages of the functional material discovery process. The left y-axis is generalized, while the right y-axis focuses on solution processed photovoltaic materials. A timeline on the bottom shows the publication date, first author and, if available, name of the platforms.*

The purpose of automation in R&D (different to automation in production) is not so much about being faster, but it is rather about guaranteeing highest data quality, best possible reproducibility, control over all parameters and the complete collection of all data in a database. Automated laboratory systems can relieve scientists from repetitive laboratory work and give them more time to design and evaluate their experiments carefully. For processes which involve manual routines, like e.g. pipetting, they offer a high reproducibility. Automated systems also make it easy to systematically collect large amounts of additional data, like timings, temperature, humidity or other conditions which a researcher often would not collect in her lab book for later evaluation (figure 1c: automated research). Yet, even with the benefits of automation, the systematic and complete screening of all parameter dimensions is not feasible in high dimensional spaces due to the large number of experiments required. Employing closed-loop inline AI optimization to guide the search for an optimum reduces the amount of experiments. The enabler to properly use AI is a systematic data collection in a machine readable format. Utilizing computational predictions allows MAPs to autonomously select which points to probe while scientists can focus on the experimental hypothesis and the inference from the collected data (figure 1c: closed-loop research). This approach makes much better use of the intellectual resources of the scientists. At the same time the reproducibility of automated processing provides the required data quality for a precise reconstruction of the complex hypersurfaces - minimizing the required samples to find an optimum. The drastic reduction in the number of experiments, as well as the scalability of the approach, are the driving force for an accelerated discovery of innovative functional materials.

The advantages of automated setups have led to the increased use of such systems in recent years. The life sciences were among the first fields to adopt equipment with features of MAPs. The field of pharmacy and drug research was revolutionized through the introduction of automation by Mitsuhide Sasaki from 1981-1992 at the Kochi Medical School [4, 5]. The field of synthetic chemistry was also among the "early adopters" of automation, with first reports on automated synthesis dating all the way back to 1966 [41]. An early overview for automated setups in the field of chemistry was already compiled by Lindsey in 1992 [42]. Highly automated setups comparable to MAPs were first introduced in the fields of systems biology with landmark systems like ADAM and EVE [43, 44], which already combined high degrees of automation, data extraction and hypothesis driven closed loop repetition for dedicated tasks like yeast or drug screening.

Figure 1d captures the evolution of automated setups in materials science over the last decade, categorizing their degree of automation and their capabilities at different stages of the functional materials development process. The figure focuses somewhat on platforms for photovoltaic materials, yet we included selected other notable robotic systems for reference and comparison. Their capabilities from chemical synthesis up to stability testing are shown on the y-axis. The left labels categorize the generalized materials development process into stages from synthesis to qualification via stress tests, which are implemented by the MAPs. Specific designations for OPV research are shown on the right side, covering the range from polymer synthesis to stability testing.

The first step in the development of new functional materials is the *chemical synthesis* (e.g. an organic semiconducting polymer). In order to process these synthesized materials, the substances must be brought into a *precursor* state (e.g. a formulated ink made from an organic polymer). If solution-based processes are used, this typically involves steps like mixing, filtering, dispersing or dissolving. The precursors are then processed into some *component* (e.g. a layer on a substrate in OPV). Depending on the field of application this component can have many manifestations. To get the desired *functionality* additional steps or materials usually are applied, e.g. deposition of an electrode grid, interlayers or protective coatings. If the functionality needed for a specific application like the power conversion for a solar cell is achieved, the temporal variation of that property can be tracked. This is usually done by a stress test, such as a light or heat induced stability quantification in photovoltaics.

By applying these categories of the materials development process we examine the development of MAPs. Our particular focus lies on advances in the automation for chemical synthesis and solution processed photovoltaics over the last decade.

**MAPs in the field of chemical synthesis**

The field of chemistry has been quite strong in the development of automated setups [42], and a considerable amount of MAPs specialize on *chemical synthesis*. *WANDA* from Chan et al. [11], first introduced in 2008, is among the earliest systems used for the development of functional materials. WANDA is a pipetting robot with a specially developed low thermal mass reactor for fully automated solution-based synthesis of colloidal nanocrystals. WANDA was utilized to perform high-throughput mapping of materials ratios and other process conditions with the purpose to optimize photoluminescence yield and peak width, upconversion luminescence and polydispersity of different nanocrystals. WANDA combined liquid handling with the ability to transport and handle vials, was built by adapting reasonably standard systems from pharmaceutical research, and required significant human interaction to run.

Another important *synthesis* system is *ARES* from Nikolaev et al. [10, 45, 46] which combined chemical vapor deposition for carbon nanotube synthesis with an in-situ raman spectroscopy for direct distinction between single-walled and multi-walled carbon nanotubes. The 4-dimensional parameter space of the nanotube growth is spanned by reaction temperature, ethene and hydrogen partial pressure and water content. While the synthesis and analysis of the nanotubes are automated, the prediction of further parameter combinations by a linear regression model was first conducted manually off-line. In 2016 the authors reported an update to their system which has since been able to perform in-line closed-loop prediction.

In the past five years the systems have become more and more sophisticated. For the *chemical synthesis* of small molecules Li et al. demonstrated a highly automated platform in 2015 [47]. They utilize a system for building-block based synthesis with standardized chemical reactions to synthesize 14 different classes of small molecules. They also discovered a commonly applicable catch-and-release purification based on a common key intermediate. The synthesis is broken down to deprotection, coupling and purification and works with a lot of commercially available building blocks. However, the synthesis path must be predefined by the user, an autonomous operation of the system is not described.

Steiner et al. in 2019 describe their *Chemputer [47, 48]*, a hardware that consists of an extensible backbone of syringe pumps and 6-way-valves to move reactants across the system. The synthesis module is abstracted to four general steps, namely reaction, work-up, isolation and purification. The abstraction is formalized to a chemical programming language, out of which the so-called *Chempiler* can generate low-level instructions for the hardware modules. For synthesis of a new molecule a retrosynthetic module generates a synthesis route, which is then carried out by the synthesis module. The *Chemputer* was validated on the *chemical synthesis* of three pharmaceutical compounds with yields and purities comparable or better than synthesis by hand.

A comparably unconventional approach for MAPs is described by Burger et al. [13]. Their *mobile robotic chemist* is composed of a robot arm mounted on top of an automated guided vehicle. It operates in a customary laboratory by driving through it from one workstation to another and using the arm to operate each station like a human would. The platform is able to operate many hours under closed-loop conditions, where new experiments are suggested by a Bayesian optimization algorithm. An optimization of photocatalysts mixtures for water splitting in a 10 dimensional parameter space is shown. The system mixes the *precursor* materials to form a *functional* photocatalyst-*component*. These catalysts were subsequently illuminated and the hydrogen evolution was measured. Five hypotheses were formulated and simultaneously tested. After 688 experiments and nearly 8 days of processing, a composition was found which showed a 6-fold higher hydrogen evolution compared to the starting conditions.

**MAPs in the field of solution processed photovoltaics**

Recently, an increasing number of reports on MAPs for photovoltaic research is seen. The platforms described so far mainly synthesize and characterize chemical compounds. In contrast to this, the following PV systems are not capable of performing complex chemical reactions. These platforms rather start by mixing different compounds and can go as far as stability testing functional solar cells.

A closed-loop synthesis platform for perovskite crystals called *ESCALATE* is described by Pendleton et al. [14] also in 2019. The focus of the work is on the orchestration software, which is an open source ontological framework for machine-readable experiment specification. *ESCALATE* comes with an abstraction layer for human interaction to simplify the initial data gathering process and facilitate the application of machine learning algorithms. It also offers the advantage that it can automatically generate a report from the available data. They demonstrate collecting a large dataset of metal-halide perovskite single crystal reactions on a commercial pipetting robot platform by preparing *precursor inks* and letting them crystallize.

Chen et al. [49] describe an automated high throughput synthesis and screening platform for mixed perovskite materials. They combined different binary compositions of iodide and bromide cations to optimize stable high-bandgap perovskite materials. *Precursor inks* were mixed by a pipetting robot and subsequently an antisolvent was added to start a precipitation process for forming polycrystalline perovskite particles. The most promising candidates were then manually processed into perovskite *thin films* and further to *solar cell* devices.

A similar process is shown by Xie et al. [49, 50] for the synthesis of nanoparticle dispersions for organic solar cell inks. An automated pipetting robot was used to dispense a donor:acceptor precursor solution into different alcohols in which the organic molecules formed nano precipitates. The *inks* were used to fabricate organic *thin-film solar cell* devices by hand, where they found ethanol as the most promising eco-friendly solvent for organic solar cell fabrication for P3HT:ICBA nanoparticle dispersions.

The above two examples show how ink preparation and optimization for solution based solar cells can be automated. Yet, the formation of thin films out of the inks is still conducted manually. The following MAPs focus on automated thin film formation out of liquid inks and go even beyond by producing complete solar cell devices and characterize them subsequently.

A closed-loop MAP for thin film formation and characterization is *Ada* by MacLeod et al. [15, 49, 50] *Ada* is a robotic platform for autonomous optimization of optical and electrical film properties. The modular platform uses a Bayesian optimization algorithm to suggest modifications of processing conditions and material compositions. The setup was demonstrated by the optimization of the hole conductivity of organic hole transporting layers. The *inks* are prepared by a pipetting channel out of stock solutions and spin coated to form *thin films* on glass substrates. The samples are then thermally annealed, photographed, optically characterized by UV-VIS spectroscopy and electrically measured by a four point probe. Subsequently, the hole mobility is calculated out of the measurement data and fed to an implementation of the Bayesian optimization algorithm *Phoenics*. The whole setup is controlled by the software *ChemOS [51]* which also interfaces to the AI for the closed-loop suggestion of the next sampling set.

Another application of the Bayesian optimization algorithm *Phoenics* is demonstrated by Langner et al. [52] They investigated two 4-dimensional parameters spaces, each created by the mixing of two organic electron donor and two electron acceptor materials. The aim of that work was to find light-stable material combinations for the photoactive layer of organic solar cells. The mixing ratios were suggested by the Bayesian optimization algorithm *Phoenics*. The suggested *inks* were mixed automatically on a pipetting robot and the resulting solutions were drop cast onto glass samples by the robot. After drying, *thin films* of the material compositions were formed. UV-VIS absorption spectra were obtained of each film. Samples were manually transferred under a metal halide lamp and illuminated for 18 hours at a light intensity of 100 mW/cm². Subsequent to this *stability test* a second absorption measurement was obtained. The integral change of the two measurements was calculated and fed back to *Phoenics* as a value for photo-induced degradation. Compared to a conducted high-throughput screening of the 4 dimensional parameter spaces, the time for finding a material combination, optimized for photo-stability was reduced by a factor of around 32.

The first, and up to now only, automated laboratory system for the manufacture and characterization of solution processed solar cells was already described by Walter et al. in 2009 [12]. *HTF-7* as they call their fabrication machine, is based on a multi-channel pipetting robot which is not only able to handle solutions and thus prepare *inks* for further processing, but also to move samples. They demonstrated a fully automated spin coating process of a hole transport layer, followed by subsequent thermal treatment and a second spin coating step in which the active photovoltaic *thin film* is spin coated. Thereafter the samples are transferred manually to an evaporation chamber for electrode deposition. The prepared functional *solar cell* samples are subsequently measured on a second automated setup, designated for optical and electrical characterization called *HTC-7*. This setup is able to automatically measure the current voltage characteristics (IV) and the external quantum efficiency (EQE) of the solar cell devices. A case study on OPV devices investigates the influence of the film thicknesses of the hole transport layer and the active layer on the power conversion efficiency. However, the platform is not limited to photovoltaic applications. In principle, all opto electronic thin film devices processed from organic solvents can be processed and characterized on these setups, which was demonstrated by showing studies on organic field effect transistors (OFET) and organic light emitting diodes (OLED).

**What comes next?**

The degree of automation and autonomy of the platforms described in the previous sections ranges from semi-automatic and partially hand-operated to setups that operate in automated closed-loop processes.

In general, the functionality of automated systems is limited to a certain area of application. All presented platforms cover a certain stage of the whole materials development process. How many stages of that process a MAP covers determines how much of the research is automatically conducted and consequently influences the research acceleration factor. The laboratory of the future goes beyond just conducting individual experiments and predicting new experimental parameters based on gathered measurement results. It is a combination of several stages served by automation and driven by algorithms that gather scientific experience across these stages from every single datapoint of all setups. This knowledge can be utilized to find to-date unknown chemical/physical interdependencies like structure-property relationships or answer other fundamental scientific questions.

On the way to this vision, the next generation of MAPs must cover a multitude of research stages, be able to capture complete process data sets, execute flexible processes, easily integrate new features through a modular design and communicate with other entities like remote AI, external databases or even other MAPs.

With *AMANDA* we are trying to follow this path. The developed system is an universal and extensible research tool for automatic and autonomous investigation of material science problems. It consists of a software framework, that provides a generic research automation system to operate MAPs and complete research laboratories. A fully operational and automated Materials Acceleration Platform for the purpose of manufacturing and characterizing lab-scale organic photovoltaic devices is included within the *AMANDA* platform. This system with the name *LineOne (L1)* consists of ca. 150 automated devices of 37 types orchestrated by the *AMANDA* software. *L1* conducts experiments on solution based thin film devices without human intervention during the process. It starts from stock solutions and cleaned substrates, is able to prepare *inks* and is capable of forming a stack of *thin films* by spin coating and thermal evaporation. Thermal treatment, optical and electronic measurements of thin films and functional *solar cells* are available as well as an accelerated aging setup to test layers and OPV devices on their *stability* with light intensities of up to 15 sun equivalents. The system offers full flexibility in the arrangement of the experiments and has a modular machine learning interface to utilize closed-loop operation with different algorithms.

# System Description – The *AMANDA* Platform

The *AMANDA* Platform is a generic research tool for distributed materials research consisting of a self developed software backbone which is capable of controlling multiple MAPs. It is able to coordinate the automated laboratory equipment of the affiliated MAPs, store all process information and measurement results in a central database and employ artificial intelligence to accelerate materials optimization. One of the controlled MAPs is *L1*. It is composed of about 150 devices of 37 different types and is designed to fabricate and characterize solution processed thin film devices like organic solar cells.

### *AMANDA* Software Framework

On the software side, *AMANDA* was designed with flexibility as a major priority and by minimizing systematic limitations as much as possible. As a consequence, the developed framework supports a large variety of equipment, with a toolset for quick integration of new devices into the system. Through this flexible approach the *AMANDA* control software is able to model a very large number of processes. Instead of hard programmed sequences, the ability to operate on so-called sequence plans is provided. These are highly customizable process recipes for the research facility which can be either arbitrarily arranged or automatically generated. They enable working with and tracking of variables as well as the usage of standard control flow operations (e.g. loops, cases, jumps) in the process. This allows the conduction of any desired researcher-defined experiment sequence within the capabilities of the given hardware.

The sequence plans are handled in the control module of the system. This module provides means of parallelization through a token based system [53] in order to make the best possible use of the equipment available and increase the average machine utilization time. Apart from parallelization *AMANDA* also has the capability of handling multiple hardware research facilities with the same software and database. This is important, especially when it comes to scalability. The hardware structure can be duplicated easily if there is an increased demand for research capacity. Likewise it might be necessary to break the workflow, e.g. to perform a particular characterization that is not possible in the line but on another setup. This feature of distributed control goes beyond just using a common control program for the facilities or a common user interface.

To make the distributed control possible, *AMANDA* intrinsically assigns every item that is processed a unique sample ID. For example, with the facility *L1* we are able to conduct complete experiments, starting at stock solutions, continuing with substrate handling and layer formation, evaporation of back electrodes and characterization of the working cells. It is possible to remove samples from this sequence, perform process steps or analytics on these samples, and return them into the sequence at a later point in time. Through the unique IDs, samples can be tracked and handled throughout the complete laboratory "landscape".

In order to implement that level of data penetration, a considerable effort was put into properly defining the database. While still striving to be as generic as possible with respect to potential recorded data, the database follows the FAIR guiding principles for scientific data management and stewardship [54]. This especially relates to the referencing, the cross-linking and traceability of data entries. In addition, the developed software by design tracks the actions with all the materials, substrates and supplies from registration upon delivery, over processing until final discarding. Having a full history on every item allows further analyses to be conducted that have yet been outside of the scope of imagination. In-process measurements give further unprecedented insights of the sample fabrication process and allow the correlation of individual process characteristics with the performance of the finished sample.

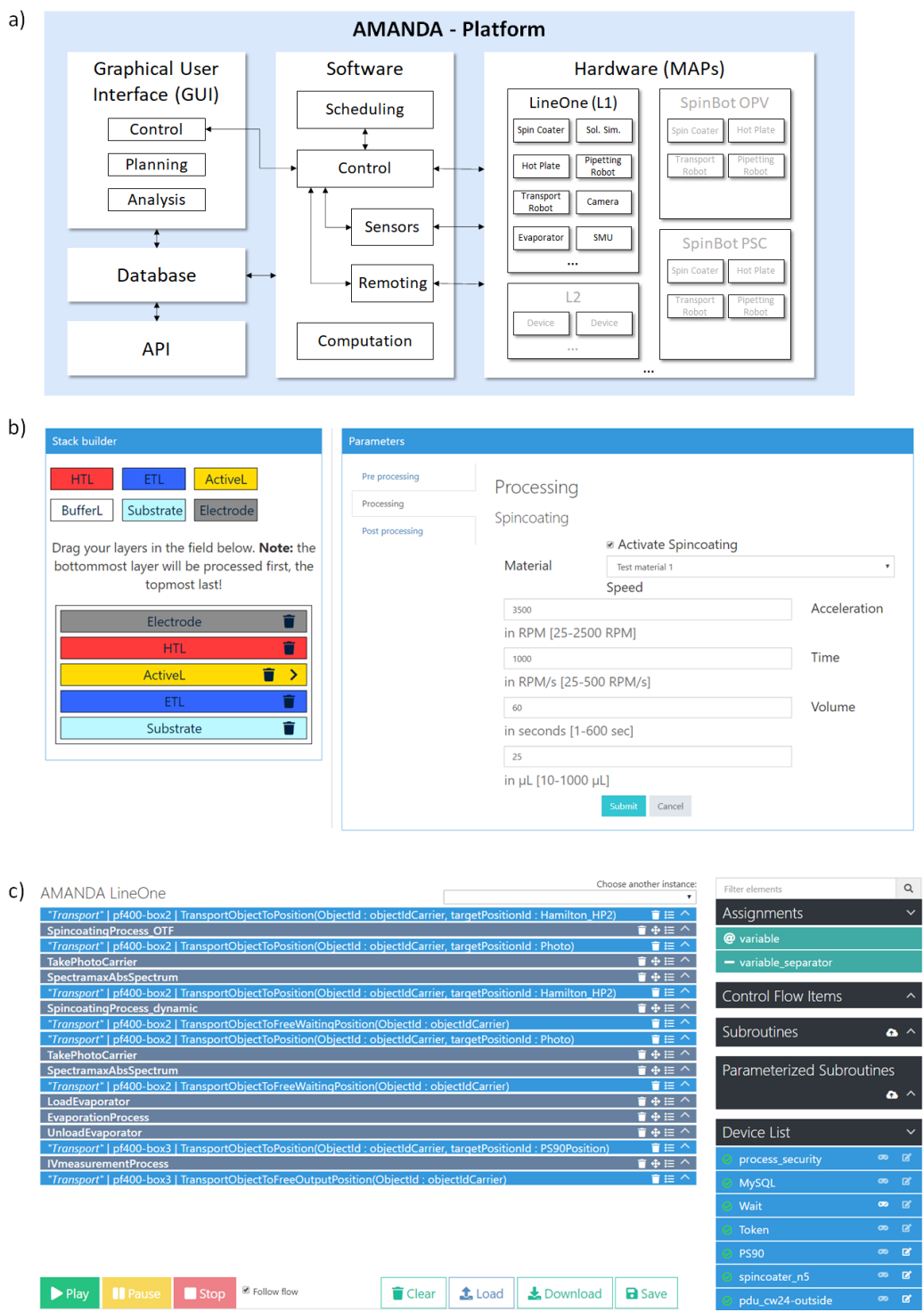

*Fig. 2: a) Overview of the Autonomous Materials and Device Application Platform AMANDA components. The left two columns represent components of the software framework, while the hardware implementations are shown on the right. Each hardware facility fulfills a specific function, e.g. L1 for the manufacturing and evaluation of opto-electronic samples. The grayed out systems are not discussed further in this publication b) Screenshot of the graphical user interface of AMANDA. Example of the 'Stack configurator' - a tool for designing an experiment by visually configuring a stack of liquid processed thin films and the*

*process parameters. c) Screenshot of the 'Sequence Plan' - A drag and drop graphical user interface for freely designing arbitrary process sequences and executing them.*

Going beyond just the automated conduction of experiments, *AMANDA* follows an Experiment-as-a-Service approach. Experiments are planned by researchers, sent as "print job" into the queue and eventually processed by the research facilities. With the introduction of abstraction layers, especially through visual configurators, researchers can focus on the experimental design rather than understanding the hardware setup. For the *L1* implementation the sequence of layers in an OPV-cell is considered as a stack. To define the process for each layer a stack configurator with adjustable parameters, including pre- and post-treatment options, allows the graphical configuration of a stack (see figure 2b) through a drag and drop approach. This stack is then translated into a sequence plan (see figure 2c) of individual process steps, broken down into machine commands, for direct execution on *AMANDA*. Automatically generated reports and interactive graphs give a quick insight into the experimental results as soon as a measurement of a sample has been conducted.

Instead of providing desktop software, an all-web-based approach was followed and the complete graphical user interface (GUI) is provided on the Internet, usable with common web browsers like Edge, Firefox or Chrome. As a consequence, researchers can plan, view and adjust their experiments from anywhere with Internet access, independent of location or terminal device.

The flexibility is also reflected by the data sources and targets for the system. The capability of interfacing with external entities, especially databases or other algorithms is designed into *AMANDA*'s core. External data sources of any kind (e.g. files, APIs, etc.) can be read in and are integrated with proper data reference and data tagging. Furthermore, a web-based API for external interfacing with e.g. experiment planners or ML algorithms is provided. The API is following a RESTful approach and is written in OpenApi 3.0 specification for quick client integration into existing systems. It exposes, according to a user permission system, access to many features like project, experiment and job management, measurement results or process logs.

**The Materials Acceleration Platform *LineOne (L1)***

The *AMANDA* orchestrated research facility *L1* consists of 3 compartments (glove boxes for future processing in inert atmosphere), each separated by controllable pneumatic doors or air locks. A picture and a schematic view of *L1* is shown in figure 3. Each glove box can be loaded from the outside into preparation areas to provide materials and disposables to the system. Three transporting robots carry substrates, solutions and disposables between different stations within the line. A flange-mounted evaporation unit is attached to glove box 3 and is automatically loaded by a special transport robot.

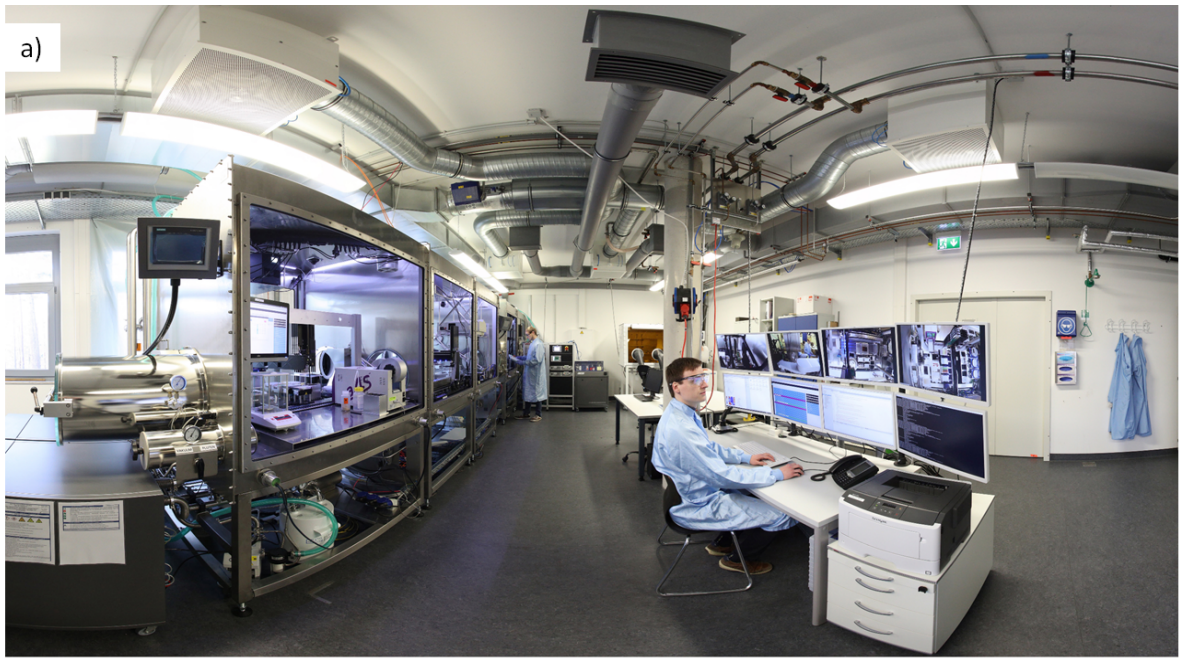
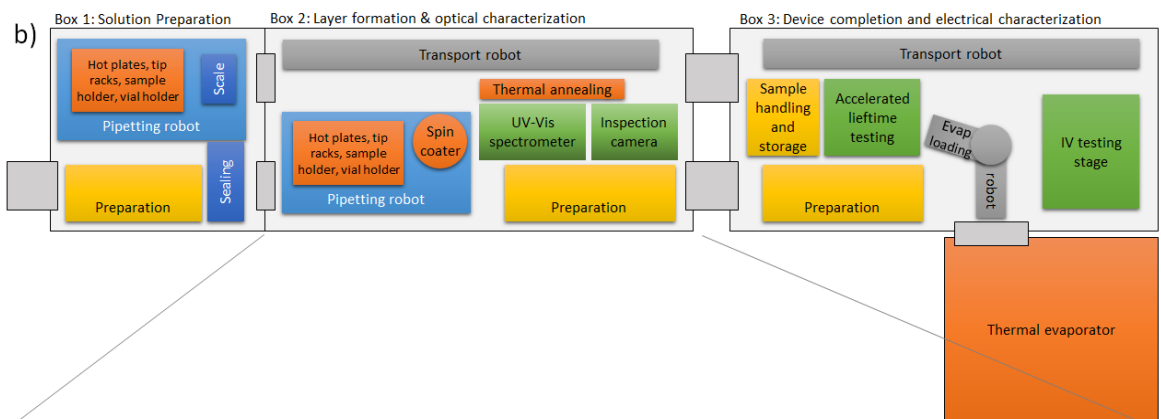
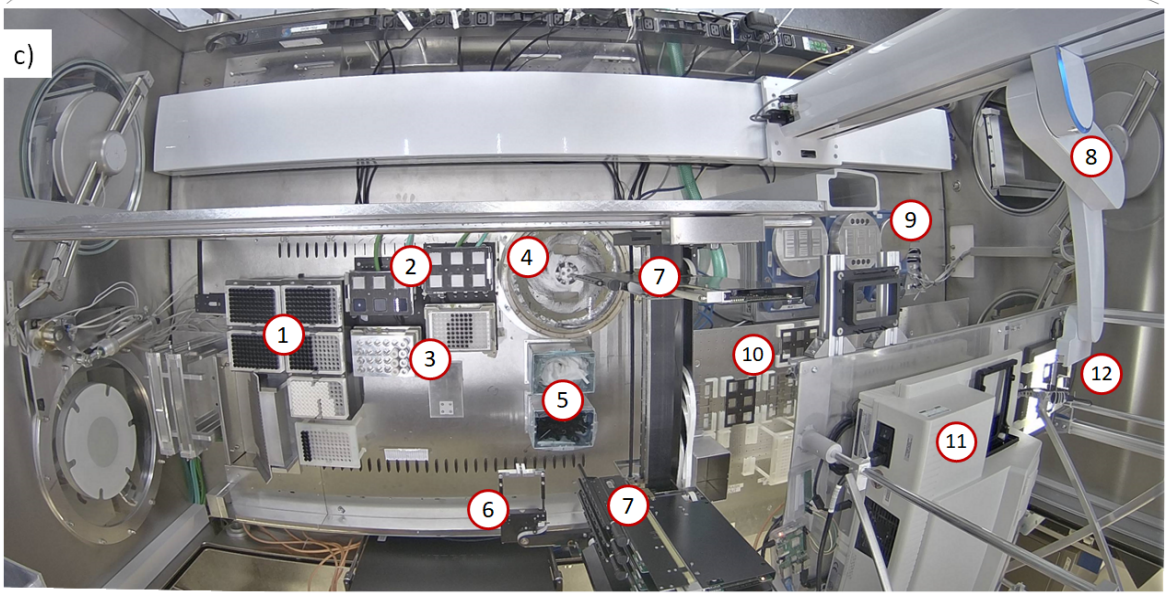

*Fig. 3: a) Overview of the AMANDA laboratory with L1 on the left side and the control desk on the right. b) Schematic view of L1 and its components. Box 1: Solution preparation;*

*including a pipetting robot, hot plates (HP), a scale and a sealing unit. Box 2: Spin coating and optical characterization; including a transport robot, a pipetting robot with hot plates, a spin coater, a UV-VIS spectrometer and an inspection camera. Box 3: Device completion and electrical characterization; including transport robots, a storage compartment, accelerated aging module, current-voltage (IV) measurement station and an evaporation unit. c) Top view of the second glove box of L1. The pipetting tips (1), substrate carriers (2), vial holders (3), spin coater (4) and trash bins (5) are located on the pipetting robots working area. The gripper (6) and pipetting channels (7) of the pipetting robot are also shown. Additionally there is a transport robot (8), hot plates for thermal annealing (9), waiting positions for substrate carriers (10), the UV-VIS-spectrometer and the inspection camera (12).*

Since most of the layers are applied with solution based techniques, the core of *L1* is the combined pipetting and spin coating station with heat shakers and hot plates as shown in figure 3b/c. Substrates are handled in batches of 6 or 8 substrates, each transported in SBS [55] sized substrate carrier frames. The used substrates are 25x25mm in size and usually contain 6 solar cells, each with an active area of 10.4 mm². Stock solutions for ink preparation are provided in v-bottom glass vials and are also transported in carriers of 24 or 32 each.

The liquid handling robot consists of 4 pipetting channels and a robotic gripper arm. The latter can transport individual samples from the substrate carrier to the spin coater and back. The pipetting channels are used to prepare coating inks from stock solutions and distribute the ink on the substrates for layer formation. We implemented two modes for spin coating (schematic in figure S1):

1. A quasi-static mode, referred to as the "on-the-fly" spin coating technique. Here, the substrate is rotated slowly (~60 rpm) while the pipette with the coating ink moves linearly in close proximity over the substrate while distributing the ink in a slow and uniform manner. When the substrate is completely covered with the solution, the spin coater accelerates to a defined target velocity.

2. A dynamic mode, where the substrate is already rotating at its target velocity and the ink is then dispensed at high speed into the center of the rotating substrate.

After ink deposition the centrifugal forces of the rotation drag the excess solution over the substrate towards their edges, creating a wet film. The film quality is influenced by many parameters like spin coating velocity, applied volume, solution properties, surface energies etc. Subsequently, the residual solvent evaporates, leaving behind a dry layer.

Another available layer application process is drop casting. A small amount of solution is dispensed directly onto a sample, leaving it there to dry under adjustable circumstances e.g. elevated temperatures of the sample.

There are options for thermal sample treatment in *L1*. Several automated hot plates enable thermal annealing of samples under various conditions like reduced atmospheric pressure and temperatures of up to 380°C. Solutions can also be treated, and are usually positioned on heat shakers to both mix and temperate them.

*L1* furthermore offers possibilities for layer characterization. A camera setup can record high resolution photographs of the samples. Those records can be used by *AMANDA*'s image analysis capability to calculate layer properties like surface coverage, homogeneity and defect density. Moreover, there is a UV-Vis spectrometer with an integrated x-y-stage for optical absorbance and photoluminescence measurements at distinct spots on the sample. As a standard procedure optical data is recorded in the areas where the solar cells will be located on the finished sample.

Layer formation on this setup is not limited to solution based methods, but can also be conducted by thermal evaporation. A two-source evaporator is integrated into the setup and accessible for the transport robots, allowing an automatically controlled deposition of up to two different materials. To define active areas for the opto-electronic cells, various shadow masks can be attached onto the samples prior to loading, allowing different geometries and sizes of the evaporated top electrode.

The layer stack can also be characterized by opto-electrical means using an LED-based class AAA solar simulator (Wavelabs Sinus 70) and several source measurement units. Current-voltage-characteristics of each cell can be recorded and are subsequently evaluated for the characteristic performance values like power conversion efficiency, short circuit current, etc. The light intensity of the solar simulator can be adjusted in order to gain an insight into the light intensity dependence of the JV-curves of the produced samples.

In order to evaluate the long-term durability of the materials used, completed solar cells can be aged in an accelerated lifetime setup. Up to 64 samples can be aged simultaneously while their temperature and the light intensity are tracked continuously. The LED light source for this characterization can reach up to 15 times the sun's light intensity.

All tasks described in this chapter can be performed fully automated by the system, in arbitrary order and without human intervention. With these tasks the *L1* hardware platform is able to conduct any solution based thin film process which involves techniques like mixing of solutions, processing layers and performing complete sample characterization.

# Mapping the OPV problem onto *AMANDA L1*

The capabilities of *L1* and the means of controlling it flexibly with the *AMANDA* software backbone allow performing systematic experiments for the evaluation of solution processed OPV systems.

In *AMANDA* all processes necessary to manufacture an OPV cell are considered a sequence of individual tasks, each of which represents a fixed, completed action with a certain set of parameters. Examples are a thermal treatment of a sample at a certain temperature for a certain duration or a spin coating task with its processing parameters. The whole experiment can then be described as an arbitrary list of tasks, which constitute a so-called sequence plan.

Figure 4 shows an example on how one full production and characterization of an OPV sample is mapped into individual tasks. The sequence plan, represented by the tasks, is stored in the database and waits as a "print job" to be triggered by an operator. *AMANDA* then informs the operator in the lab to load fresh substrates and stock solutions into *L1*. The production process starts with a reference measurement of the blank substrate consisting of a photograph and an absorbance measurement. In parallel to that characterization, the solution for the first layer is prepared. The desired amounts of the stock solutions are each aspirated into one of the pipetting channels and dispensed into a common, small container. The containment is shaken in order to achieve a good blend before the mixture is used in the process. Subsequently, the first layer is formed by spin coating and in predefined areas removed again with a cotton swab. This cleansing is required to enable an electrical contact to the bottom transparent electrode layer for the later conducted electrical characterization. For this purpose, a cotton swab is picked up by a pipetting channel and wiped over the substrate short before the layer is thermally annealed. This procedure (characterization and solution mixing, spin coating, wiping, annealing) is repeated until all solution based layers are formed. Another optical characterization is conducted before the final top electrode layer is applied by thermal evaporation. Finally, a current-voltage measurement of the completed sample is taken in dark and under AM1.5 illumination by utilizing a solar simulator. A stability measurement can optionally be performed on the solar cells. For this purpose, the samples are transferred to the module for accelerated aging and exposed to LED light of up to 15 suns. This testing can be interrupted by IV measurements in given intervals based on illumination exposure, temperature over time or fixed timings to gain insight into the changing performance behavior. Upon completion of the whole process, the sample is automatically transferred into a storage unit.

The above description shows the process for one individual sample. In *L1* and in other systems on the *AMANDA* Platform we work in batches of 6 or 8 samples to allow parallelization and increase throughput.

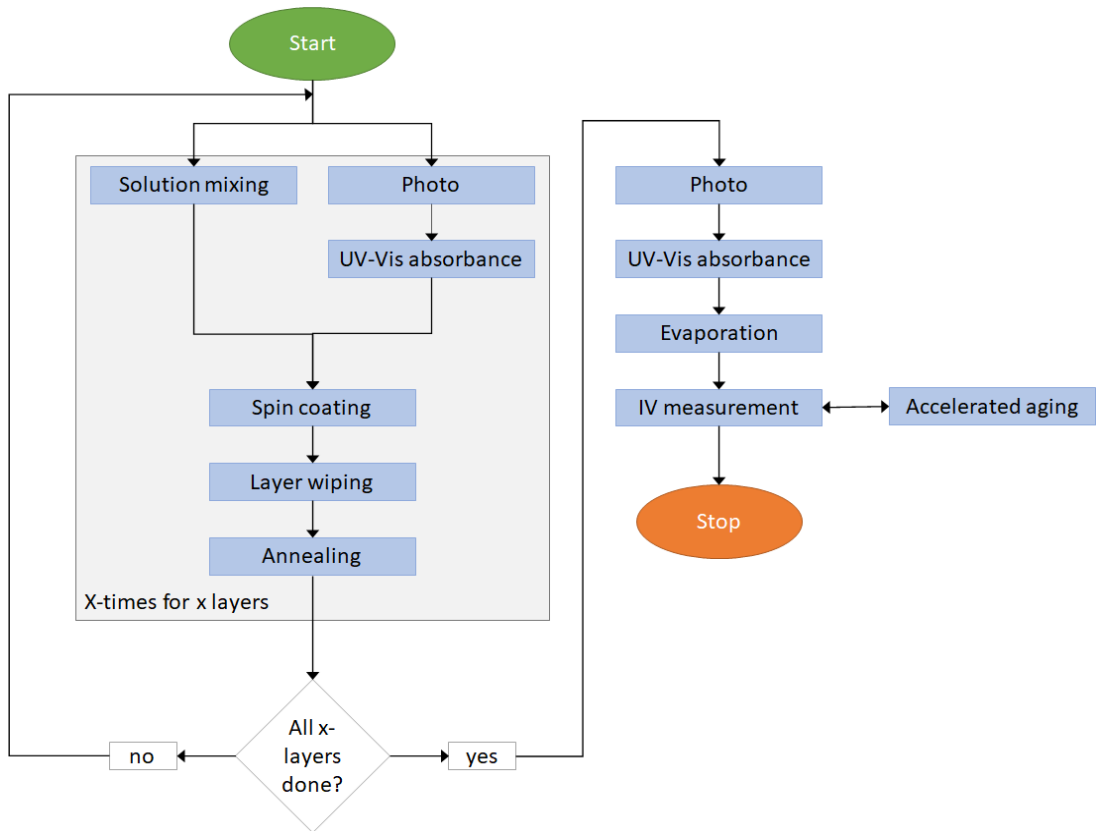

*Fig. 4: Process sequence for an OPV sample fabrication process on L1, divided into individual processing tasks.*

# Experimental Results

To demonstrate the capabilities of producing state-of-the-art OPV cells, we present experimental results from the optimization of the PM6:Y6 bulk heterojunction system in air performed on *AMANDA L1*. Three use cases of optimization are described. A concentration study demonstrates the ability of mixing solutions and feeding them into the sample production process. In a spin coating study the effect of the deposited solution volume on the layer quality and sample performance is investigated. Finally a long term reproducibility study across many processing runs and processing days is performed that shows a small distribution of measured solar cell performance.

**Fully automated OPV process - concentration study**

For the study of the effect of active layer thickness on the device performance, automated experiments were conducted on *L1*. These were based on the flow chart shown in figure 4 in order to investigate the dependency of the PM6:Y6 solar cell performance on the thickness of the active layer. *L1* prepared different concentrations of PM6:Y6 solution in chloroform (CF), varying the solid content. For this purpose three stock solutions, based on chloroform as the main solvent, and 1-chloronaphthalene (CN) as additive (0.5%), were prepared manually in vials and fed into the system. A PM6 solution, a Y6 solution and a vial of CF:CN solvent mixture. These solutions were automatically mixed, according to the schematic in figure 5b. While the PM6:Y6 ratio was kept constant to 1:1.2 in this experiment, the total solid content of the solution was varied. The prepared solutions were subsequently spin coated and characterized. The results of this experiment are shown in figure 5c.

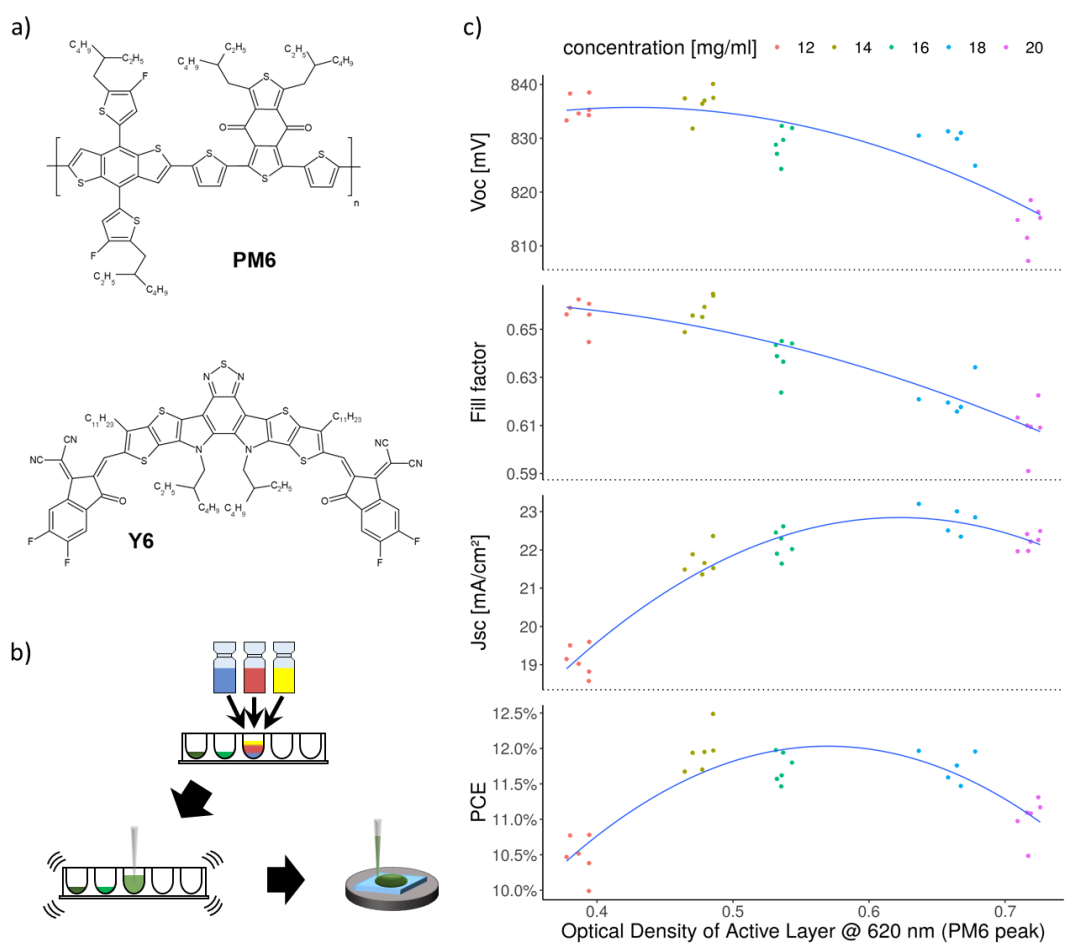

*Fig. 5: a) Chemical structure of PM6 and Y6. b) Inline mixing of spin coating solution as used in this experiment: 3 stock solutions are provided to the pipetting robot; the solutions are deposited with 3 individual pipetting steps together in one well of a well plate with ratios according to the experimental plan; the well plate is shaken to blend the solutions; the liquid handling robot aspirates the solution from the well and dispenses it in the spin coating process. c) Concentration study for OPV performance of PM6:Y6 solar cells plotted as a function of the optical density of the active layer. The blue line is a locally estimated scatterplot smoothing curve (LOESS)[56].*

The optical density (OD) was extracted from absorbance measurements of the sample before and after active layer spin coating. Since there are 6 independent solar cell devices on each sample, the OD values are measured for each of the 6 positions on the substrate where the active solar cell areas will be located. The absolute difference of the OD before and after spin coating was calculated at the PM6 absorption peak wavelength of 620 nm. The OD values and the solar cell performance values are plotted against each other in figure 5c for each of the six cells on the sample. The plot gives an insight into the layer thickness dependence of the solar cell performance.

As anticipated, the fill factor (*FF*) and the open circuit voltage ($V_{OC}$) decreases as the active layer film thickness increases. In contrast to that, the short circuit current ($J_{sc}$) drops for thinner layers. For thicker active layers the $J_{sc}$ rises until it reaches its maximum at an OD of around 0.65. The power conversion efficiency (*PCE*) is a combination of the previous values and shows a maximum at an OD of around 0.55.

First data of the automated experimentation on *L1* is presented in this study. The solution for the active layer was mixed inline by the liquid handling robots, short before spin coating it onto the sample. Thus, we were able to examine the dependency of the layer thickness with the solar cell performance. The optimum concentration of solid content was found around 16.5 mg/ml. For this procedure only the stock solutions were prepared by hand as the liquid handling robot created the appropriate mixture. With this generic approach, we are able to continuously probe the complete compositional space automatically, even for very complex, high-dimensional problems.

**Spin coating parameter study - volume**

To explore the possibility of minimizing material consumption we varied the amount of solution deposited on the sample to form the active layer during a dynamic spin coating process. In our manual routines we tend to use a solution volume ($V_{spin}$) of 50 µl for our 25 x 25 mm substrates. For this experiment, we used 50 µl as a starting point and reduced it down to 5 µl. A set of pictures taken of the active layer films while systematically decreasing the $V_{spin}$ is shown in figure 6a. For volumes at or below 15 µl one can see clearly that the substrate surface is not fully covered anymore with the wet film, which leads to torn films towards the outside of the substrates. This behavior is expected and can be easily explained by the insufficient amount of applied material.

The layer quality however seems to decrease again for higher volumes of 30 µl and above as the films tend to be increasingly inhomogeneous and show features that resemble drying stains. These artifacts in all likelihood result from too much solution in those particular spots. This excess solution is not pushed off the substrate fast enough by centrifugal forces and dries inhomogeneously on the sample surface. To avoid this behavior, the removal rate of the excess solution has to be higher than the rate of the solution's solvent evaporation.

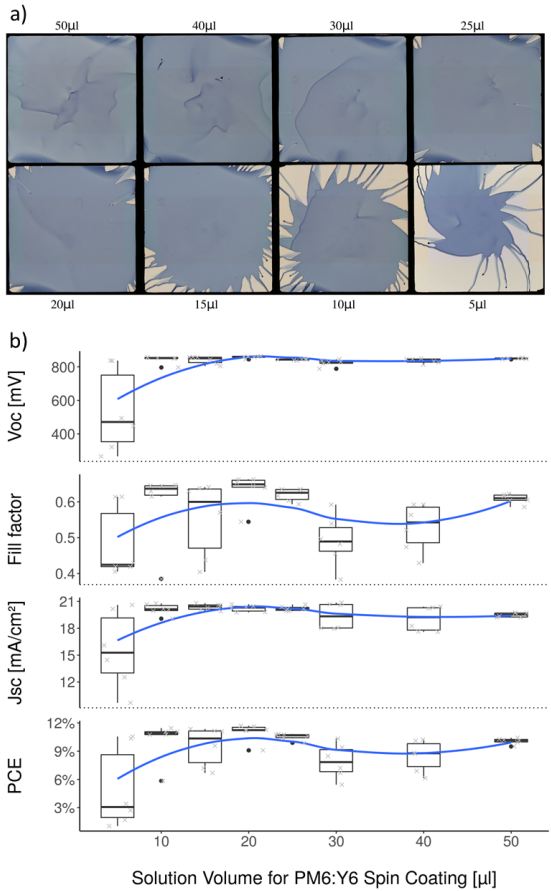

*Fig. 6: a) Photographs of PM6:Y6 layers spin coated automatically on AMANDA L1 while varying $V_{spin}$ from 5 µl to 50 µl. b) Solar cell performance of the samples shown in (a), plotted over $V_{spin}$.*

Adjusting the parameters, e.g. solvent, solvent additives, liquid volume, spin coating velocity, etc. can help to reach an optimal, homogeneous film as well as having an effect on the composition and morphology. In this experiment, we used chloroform which has a very high evaporation rate due to its low boiling point. Since the morphology formation of PM6:Y6 layers is strongly dependent on the choice of the processing solvent, a change to high boiling solvents is detrimental [57]. The control of the layer thickness is therefore made possible by changing the spin coating velocity, the solid content and the amount of used solvent. Nevertheless those parameters can also influence the drying kinetics and thus lead to a different microstructure of the resulting film. It is therefore important that those parameters can be adjusted individually when trying to find optimal process conditions.

Altogether in our experiments, the layers with the best visual appearance are formed with a $V_{spin}$ of 20 µl and 25 µl. In figure 6b the solar cell performance is plotted as a function of the $V_{spin}$ and a similar behavior as compared to the photographs can be observed. For 5 µl, the performance drops dramatically, as the active layer film coverage over all the solar cells of a sample is not given. For 30 µl and 40 µl the performance varies especially in fill factor and short circuit current, which can be attributed to poor layer homogeneity. Overall the best results were achieved with volumes of 20 and 25 μl. This is half of the amount used in our manual processes and shows *L1* can decrease the necessary material quantities per substrate by 50%. As a consequence, the automated system can reduce the material consumption and perform twice the amount of experiments with the same amount of given material.

**Demonstrating the reproducibility of *L1***

Each processing day we process at least one PM6:Y6 reference sample to control the reproducibility of the system. The process parameters of these control-samples are constant, although they are processed on different days with changes in ambient conditions, base solution and time. The diagram in figure 7, is composed of the data acquired from such reference samples over a three-month observation period and 19 different experimental runs. On each of those 19 reference samples there are 6 solar cells. The cells are considered functional if an open circuit voltage ($V_{OC}$) over 0.3 V and a fill factor (*FF*) over 0.4 were reached. This condition was achieved by 105 of the 114 solar cells, leading to a rate of 92% functional solar cells.

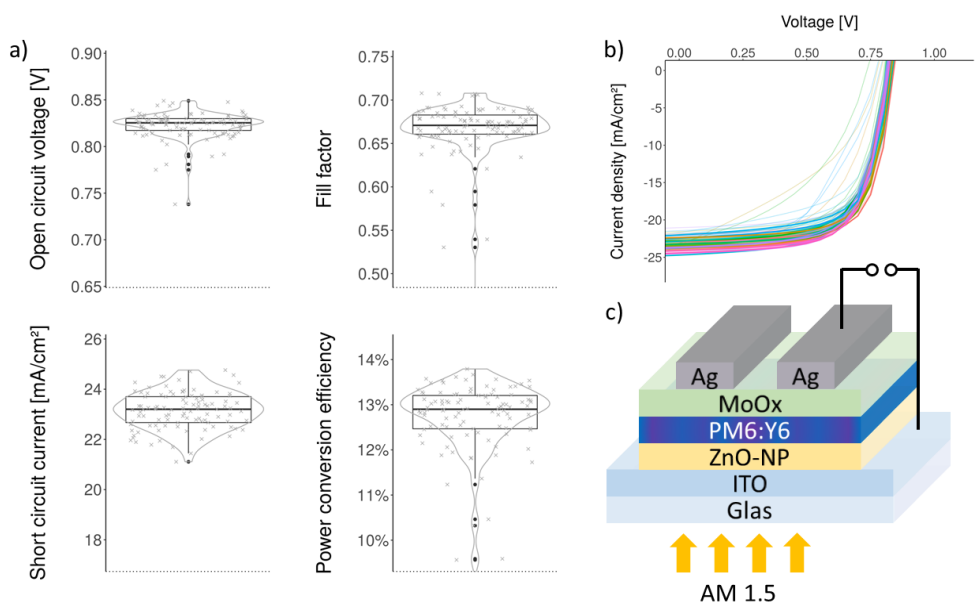

*Fig. 7: a) Distribution of the solar cell performance of PM6:Y6 reference samples from 19 separate experiments over a 3-months period. Statistics are shown as box plots with a maximum whisker length of 1.5 times interquartile range. Additionally the data points of the single solar cells are shown as a scatter plot, and their distribution as violin plot. B) JV-Curves of the 19 samples, each with 6 solar cells. The most efficient solar cell of each sample is plotted bold. c) Architecture of the solar cells used in this publication. Layer stack is: Glass, indium tin oxide, zinc oxide nanoparticles, active layer of PM6:Y6, molybdenum oxide and silver.*

The photovoltaic median performance and the interquartile range of the functional solar cells is shown in table 1. The open circuit voltage ($V_{OC}$) shows a very small interquartile range of 0.013 V with a median value of 0.825 V. The median(interquartile range) of the short circuit current ($J_{SC}$), fill factor (*FF*) and the power conversion efficiency (*PCE*) are 23,2(1,03) mA/cm², 0.671(0.022) and 12.9(0.74)% respectively.

The $V_{OC}$ value of these samples is equal to the ones in the publication of Yuan et al [25]. However short circuit current (~-2 mA/cm²), fill factor (~-0.1) and thus the power conversion efficiency (~-2.8%) are below the hero performances reported in the literature. In contrast to the solar cells of Yuan et al., the solar cell areas described in this publication are 2.5 times as large though. In addition, the samples were completely processed under ambient conditions in the present work.

*Table 1: Maximum, Median and interquartile range values for PM6:Y6 organic solar cells, processed completely automated on AMANDA L1. The shown samples were produced with the same parameter settings in 19 separate experiments, over a 3-months period.*

|  | $V_{OC}$ [V] | $J_{SC}$ [mA/cm²] | FF | PCE [%] |
| --- | --- | --- | --- | --- |
| Maximum | 0.849 | 24.8 | 0.708 | 13.79 |
| Median | 0.825 | 23.2 | 0.671 | 12.9 |
| interquartile range | 0.013 | 1.03 | 0.022 | 0.74 |

Compared to other PM6:Y6 devices processed under ambient conditions, the solar cells fabricated in this publication are also quite comparable. Sun et al. reported a maximum efficiency of blade-coated PM6:Y6 bulk-heterojunction solar cells of 15.37% for small devices of 0.04 cm² area and 14.01% for larger area devices with 1 cm² active area [37].

## AI based performance and stability prediction on LineOne

An application example outside the scope of this publication was presented by Du et al. where *AMANDA* and *LineOne* were used to investigate the efficiency and stability of PM6:Y6 solar cells as a function of different process parameters. During the production of the solar cells the absorption of each active layer was measured. After the evaporation of electrodes, the IV characteristic of all samples were obtained and a portion of the samples were subsequently aged for 50 hours at one sun. The data from the absorption spectra of the active layer was used to deconvolute the spectrum into single features, representing different properties of the active layer blend. Based on these features a Gaussian Process Regression (GPR) model was trained to predict performance and stability. They showed that the GPR model was able to predict the device performance (especially $V_{OC}$) and the photostability with high accuracy. From the model the authors identified the molecular ordering as key for the optimization of the produced solar cells. Thin active layers with medium annealing temperatures were found to be beneficial for both initial performance and photostability of the investigated material system [58].

# Challenges and Opportunities of AMANDA and the laboratory of the future

Developing a fully automated research facility like *L1* is a challenging task in many aspects. In order to start a process with solutions and conclude with fully evaluated solar cells, the whole process chain has to be broken down into many subtasks that need to be automated and coordinated. Every action usually performed by an operator needs to be mapped onto an automated system with rather limited degrees of freedom.

While not obvious at first hand, the biggest challenges for integrating an OPV process into *L1* were rather the small things that usually go unnoticed. Good examples for this are trivial actions like opening up a screw capped vial, wiping away a stripe of a 200 nm thick layer of polymer or even just putting a shadow mask onto some substrates and turning that stack around. Those tasks are straightforward for humans with 27 joints [18, 19] in one hand alone. But for a single 6 axis robot with a binary-state open/close gripper it is not simple at all. As we implemented the OPV fabrication process on *L1* we broke down the whole process into 50 single tasks which can be executed natively by one of the integrated devices (see table S2). However, for several tasks we adapted and validated alternative approaches, or had to devise attachments and workarounds. For the opening of vials, we exchanged screw caps for rubber corks, which can be picked up by one of the pipetting channels with a special adapter (figure S2b). In a similar fashion we also designed adapters that are fitted with cotton swabs for the pipetting channels in order to remove stripes of the coated polymer layers (figure S2a).

A further challenge we faced was the integration and automation of equipment originally designed for regular laboratory use. Common laboratory machines are typically designed for manual operation and are often not suited for the use in automated systems. Apart from transport of samples between and the placement into the equipment the hardware interfaces and software communication protocols needed special attention. *AMANDA* controls about 150 single devices of 37 different types in *L1*. To interface with that variety of hardware the remote connections needed to be converted to a common standard. In *AMANDA*'s case the ethernet network was selected as communication standard, both for scalability as well as the centralized server based control from a Linux environment.

Since all the integrated devices were selected by their functionality for the process and not on the availability of their documentation, a significant effort has gone into the proper implementation of communication protocols. As a result a vast library of basic implementations for nearly any connection type is part of the framework that allows very quick integration of new devices at this point. However, a small number of vendors have proven to be rather unwilling to share the specifications of their devices which required the reverse engineering of their protocols and devices. We strongly advocate the use of a common standard in laboratory equipment, one approach has been shown for example with the SiLA standard [59].

Despite these challenges the advantages of an automated and integrated system make the effort worthwhile. The complete automation improves the process reproducibility as well as the quality and especially the completeness of the acquired data for each sample. In addition, a generic toolkit like *AMANDA*, which has now proven its capabilities on OPV, can very easily be adapted to tackle questions in other fields of materials science like e.g. quantum dots, perovskite solar cells, thermoelectrics, catalysts, batteries or fuel cells. The comprehensive, structured collection of data with unique identifiers in machine readable formats generated by MAPs like *L1* or other *AMANDA* facilities is the foundation for deploying artificial intelligence to solve materials science questions. This enables the efficient (multi-objective-) optimum search in vast high dimensional spaces of materials composition and process conditions. This approach is likely to significantly accelerate the discovery and commercialization of new functional materials in the future.

A further important aspect for the acceleration of materials science is parallelization in conjunction with connectivity. The *AMANDA* software backbone allows the simultaneous control of multiple hardware research facilities and respectively various hardware devices which can be operated either in parallel, or be combined into a larger sequence. This makes it possible to create virtually connected cloud laboratories and cross platform integration of data, providing scalability for addressing complex and important materials science problems, which has never been available before. The flexible, web-based process design and the experiment-as-a-service approach of *AMANDA* provides full process control and data access for scientists, independently of their actual location opening a path to truly distributed research. A specialized facility like *L1* in the *AMANDA* environment could allow researchers, which otherwise would never have access to such research equipment, to test their ideas on a service level, extending an approach that has been commonly used for computational investigations on supercomputers into materials science facilities. As such it provides scalability with regards to the utilization of intellectual potential.

A further property the automation approach shares with computing is scalability of processing resources. Training a skilled operator on complex tasks takes years. However, when an automated facility exists it is possible to simply reproduce the facility in order to double the throughput while still providing exactly the same quality of operation. Apart from that optimization of existing facilities also has a large potential. In our implementation of *L1* there are still significant capacity increases possible through the implementation of further parallelization in *L1*. While current "process runs" on *L1* typically encompass 6 batches of 6 samples each in a typical workday, which sums up to 36 samples. Since every sample is divided in 6 parts and each part contains an independent solar cell device, this estimates to 216 solar cells. However, this does not reflect the potential throughput of the facility. With a more sophisticated scheduling algorithm for parallel processed batches *L1* would be able to process 9 batches on a regular work day of 8 hours. The batch size can also be increased to 8 samples, that would make it possible to process 72 samples (≙ 432 solar cells) in a single shift on a normal working day (compare figure S3). In continuous operation up to 34 batches could be processed in 24 hours, which corresponds to 272 samples (≙ 1632 solar cells). All this could already be achieved with minimal hardware changes. In a sequential automated process it is always the process with the longest cycle time that limits the throughput. In the case of *L1*, the evaporator is the limiting factor at the moment. With a cycle time of 42 minutes per batch a maximum of 34 evaporations per day can be achieved. To overcome this bottleneck, multiple batches would need to be evaporated at the same time, giving *L1* the ability to roughly double its throughput to ca. 500 samples with ca. 3000 devices per day. To go beyond this, the setup would need to be reproduced, which is much simpler and faster than the initial development, since from day one it would benefit from all the "knowledge" acquired with the previous setups.

# Conclusion

Automation and artificial intelligence enhanced discovery and qualification of new materials plays an increasingly important role in materials science. In this publication we reviewed the development of Materials Acceleration Platforms over the past decade. Using a generalized classification, we compared selected MAPs and examined the degree of automation of the individual stages of material development. We concluded that interconnected, multi-stage MAPs, reinforced with artificial intelligence are the way to the automated lab of the future. As one step towards this goal we presented the *AMANDA* platform, a framework for autonomous laboratory research, and the hardware facility *L1*, a full-scale Materials Acceleration Platform for generalized and flexible materials research. *AMANDA* together with the hardware implementation *L1* has reached a maturity, where fully automated processing of organic solar cell devices is conducted on a daily basis. The software backbone of *AMANDA* follows the experiment-as-a-service approach that abstracts design of experiment from automated execution and allows researchers to focus on data and evaluation. We discussed the large potential of flexible automation for science and demonstrated how to transfer a complex laboratory process onto a robotic platform. The functionality of *L1* was demonstrated on the hand of automatically manufacturing solar cells with the PM6:Y6 material system, achieving efficiencies for ambient atmosphere processing of up to 13.7% under AM1.5 illumination. The precision of the platform was used to investigate the influence on performance of different parameters that are difficult to access with manual lab work. Specifically the effect of semiconductor concentration in the solution used for spin coating, as well as the effect of the solution volume deposited for spin coating, were investigated. Optimal process parameters could easily be extracted from the results. Furthermore, the reproducibility of this system is demonstrated with repeated production runs of the PM6:Y6 material system, showing an interquartile range of 0.74% in *PCE* for 108 individual cells on 19 substrates which were processed as reference samples in 19 individual processing runs performed over the course of 3 months.

*AMANDA* together with the Materials Acceleration Platform *L1* is a work in progress, yet already now demonstrate the capabilities of such systems. Despite the fact that they are able to perform all the functions to manufacture lab scale solution-based solar cells automatically, new features are still being added, and the production capacity is continuously increased. Recently, a stability screening module was integrated in *L1* which is capable of accelerated aging of up to 64 samples in parallel at 15 suns max. Its integration allows the determination of efficiency potential, processing parameters and stability for a new material in one process. In general we expect that *AMANDA* will be capable of generating enough devices and data for AI algorithms to understand the underlying model of properties and processing conditions of a new material system within one to two processing days. However, to further speed up this process, we believe that the data for the AI computations must be extended beyond the current experimental loop and include the knowledge from other (previous) experiments as well. By finding similarities in the chemical composition of the material systems we are looking to bias the prediction model to quicker converge towards optima.

A further step for *AMANDA* and *L1* will be the completion of the discussed materials development process by including the synthesis stage. We believe that by synthesizing, processing and long-term evaluating materials in-line in one MAP, screening of their structure-property relationship will be drastically accelerated.

We believe that with MAPs a paradigm shift in materials research is imminent. Automation and AI will change the way how research is conducted over the next decade. Driven by systems like *AMANDA* with hardware research facilities like *L1*, integrated Materials Acceleration Platforms provide researchers with unprecedented means of experimentation and challenge them to think in new ways within the laboratory of the future. This has a major potential to revolutionize materials science, accelerate the discovery of new functional materials and decrease the time to market -- in the best sense of the materials genome initiative (MGI).

# Supplementary Material

See the supplementary material for more information about process tasks, used devices and a more comprehensive view of parallel execution of multiple batches on L1.

# Data Availability

# Acknowledgement

The authors want to thank the Deutsche Forschungsgemeinschaft (DFG) for financial support in scope of the DFG INST 90/917-1 FUGG and DFG BR 4031/13-1. We also gratefully acknowledge the grants "ELF-PV - Design and development of solution processed functional materials for the next generations of PV technologies" (No. 44-6521a/20/4) and "Solar Factory of the Future" (FKZ 20.2-3410.5-4-5) by the Bavarian State Government.

# Conflict of Interest

The authors declare no conflict of interest.

# Appendix: Experimental

All processing steps were performed in ambient atmosphere, except for the preparation of PM6:Y6 stock solutions and thermal annealing of active layer for the reproducibility runs. The organic solar cells fabricated for this work were deposited on indium tin oxide covered glass substrates with a sheet resistance of 7-9 Ω/sq. Before usage the substrates were cleaned in an ultrasonic bath for 10 minutes in purified water, acetone and 2-propanol respectively. For all solar cells an inverted architecture was used with the first layer comprised of zinc oxide nanoparticles (ZnO-NP), spin coated automatically on *L1* with 3000 rpm, in the quasi-static on-the-fly spin coating mode. The solution was purchased from Avantama and was treated prior to usage with an ultrasonic horn for 60 seconds before filtering with a 0.2 µm PTFE syringe filter. The ZnO-NP layer was post treated at 200°C for 5 minutes. After optical characterization of that first layer, the active layer was deposited. Therefore a chloroform stock solution was prepared with 99.5 vol% chloroform and 0.5 vol% of 1-chloronaphthalene. PM6 and Y6 purchased from Derthon were diluted in parts of the chloroform stock solution respectively, forming two additional stock solutions with a solid content of 20 mg/ml each. These three solutions were mixed by the pipetting robot before usage. The ratio of PM6 and Y6 was 1:1.2 for all shown devices, the total solid content was varied between 10 and 20 mg/ml (16 mg/ml for the reference samples). The active layer was spin coated in dynamic spin coating mode on *L1* at a rotation velocity of 3500 rpm. The used solution amount was varied between 5 µl and 50 µl. The pipette to substrate distance during ink dispensation was fixed to 50 mm. A dispense speed of 500 µl/s was used. The active layer was thermally annealed at 100°C for 10 minutes and optically characterized after annealing. Then the samples were transported to the evaporator. A shadow mask was applied to the substrates leading to an active area of the solar cells of 10.4 mm². The substrates were loaded into the evaporation chamber and molybdenum oxide and silver were evaporated subsequently with a waiting time of 2 minutes in between the two evaporation processes. The molybdenum oxide was evaporated with a rate of 0.1 Å/s for the first 2 nm, then with 0.2 Å/s until a final film thickness of 10 nm was reached. For silver an initial rate of 0.5 Å/s was used for the first 10 nm, after that the rate was increased to 3 Å/s until a total film thickness of 100 nm was reached. The solar cell performance was measured automatically in *L1* under an AM1.5 spectrum provided by a Newport LSH-7320 class ABA LED solar simulator, and additionally measured by a Wavelabs Sinus-70 class AAA LED solar simulator manually.

# Supporting Information

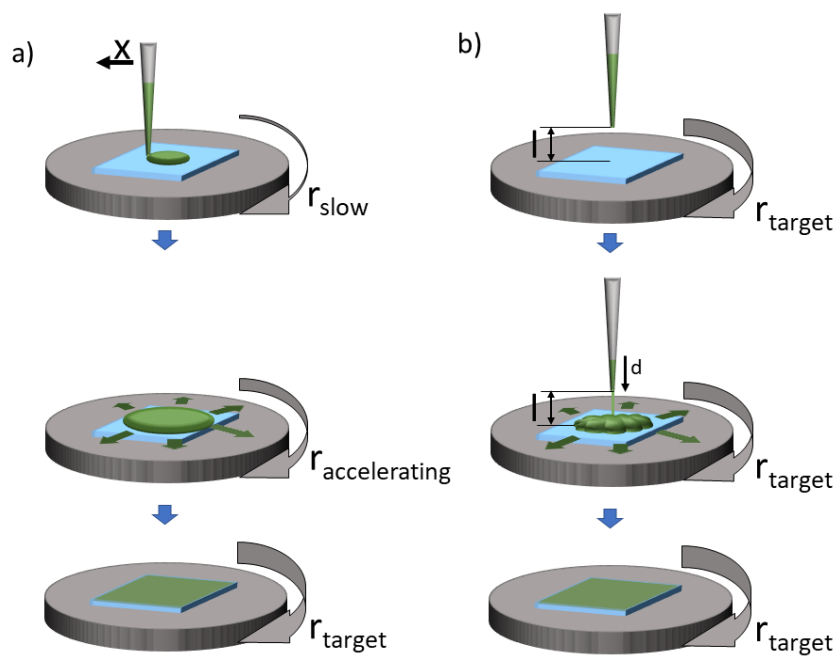

*Fig. S1: The two spin coating modes available on L1: a) quasi static mode (on-the-fly) were the solution is distributed onto the substrate with dispense speed d, while the tip moves linearly over the substrates surface with speed x and the spin coater rotates slowly with velocity $r_{slow}$. The substrate then accelerates to the target velocity $r_{target}$. b) dynamic mode. The tip is in distance l over the substrate center, the spin coater is rotating at target velocity $r_{target}$ while the solution is dispensed with dispense speed d.*

*Table S1: List of all devices in L1 controlled by AMANDA.*

| Amount | Category | Device |
|---|---|---|
| 5 | Infrastructure | Remote Controllable Power Distribution Unit |
| 2 | Robot | Liquid Handling Robot (Hamilton Starlet) |
| 2 | Robot | SCARA Robot |
| 8 | Device | Heatshaker |
| 1 | Device | Heatshaker with Gripping Mechanism |
| 1 | Virtual | WaitDevice |

| 1 | Virtual | MySQLDevice |
|---|---|---|
| 1 | Virtual | TestDevice |
| 1 | Virtual | TokenDevice |
| 1 | Virtual | MachineLearningDevice |
| 1 | Measurement | Industrial Grade Camera |
| 1 | Measurement | UV VIS Spectrometer (Molecular Devices Spectramax M2e) |
| 1 | Device | Solar Simulator (Wavelabs Sinus-70) |
| 3 | Device | Magnetic Stirrer with Heating |
| 2 | Measurement | Source Measurement Unit |
| 1 | Robot | Rotation-Centric 6 Axis Robot |
| 2 | Measurement | Scale |
| 1 | Film Application | Doctor Blade |
| 3 | Film Application | Spin Coater |
| 1 | Device | Heat Sealer |
| 13 | Infrastructure | Surveillance Cameras |
| 1 | Device | Photo Light Table |
| 4 | Sensor | Raspberry Pi + Sensors |
| 8 | Connectors | RS232 to Ethernet Adapters |
| 7 | Infrastructure | Gigabit Network Switches |
| 3 | Infrastructure | Lights inside Glovebox |
| 11 | Infrastructure | Network controllable relay |
| 1 | Infrastructure | Signal Post |
| 2 | Device | Hot Plate (in Sluice) |
| 2 | Device | Sluice |
| 7 | Infrastructure | Airlock Doors |
| 1 | Device | XYZ-Stage |
| 1 | Film Application | Evaporator (Leybold Univex 250) |
| 1 | Film Application | Evaporator Film Controller |
| 1 | Measurement | Measurement SMU Switch |
| 48 | Stability | LED Light Controller |
| 3 | Infrastructure | Radiator |

*Table S2: List of all processing tasks necessary to fabricate and characterize organic solar cells in AMANDA L1, ordered after their first occurence in the standard process (compare figure 4). Some of the steps appear multiple times at different stages during the process.*

| | Task | Equipment Involved | parameters | Task description |
|---|---|---|---|---|
| 1 | Register substrates and substrate carrier | *AMANDA* Control | substrate type | Generate a virtual entity for substrate carrier and its containing samples and assigns sample IDs |
| 2 | Pick up substrate carrier | SCARA Transport Robot | carrier ID | Pick up a substrate carrier, identified by its ID |
| 3 | Put down substrate carrier | SCARA Transport Robot | target position | Put the substrate carrier to the target position |
| 4 | Pick up vial holder | SCARA Transport Robot | vial holder ID | Pick up a certain vial holder, identified by its ID |

| 5 | Put down vial holder | SCARA Transport Robot | target position | Put the vial holder to the target position |
|---|---|---|---|---|
| 6 | Turn on heat shaker | Hot plate | target Temperature, target shaker intensity | Set target temperature and shaker intensity of the hot plate and starts heating/shaking |
| 7 | Pick up single substrate | Pipetting Robot | substrate ID | Pick up a substrate, identified by its ID |
| 8 | Put down single substrate | Pipetting Robot | target position | Put substrate to target position |
| 9 | Pick up pipetting tip | Pipetting Robot | tip size | Pick up next available pipetting tip of tip size |
| 10 | Decap vial | Pipetting Robot | target vial | Open the target vial |
| 11 | Aspirate solution | Pipetting Robot | target vial, volume, aspiration speed, mixing parameters | Aspiration of solution into the pipetting tip with the given parameters |
| 12 | Dispense solution in vial | Pipetting Robot | target vial, volume, dispense speed | Dispensing of Solution into vial, e.g. for Mixing of multiple solutions |
| 13 | Start spin coater | Spin coater | target velocity, acceleration, duration | Start the rotational movement of the spin coater with the given parameters |
| 14 | Dispense solution onto substrate | Pipetting Robot | substrate ID, volume, dispense speed, distance from substrate, distance from substrate center | Dispense solution onto a substrate, the substrate can be still (=drop casting), or rotated by the spin coater (= dynamic spin coating) |
| 15 | Dispense solution onto substrate -- on-the-fly | Pipetting Robot | substrate ID, volume, dispense speed, distance from substrate, movement | Dispensing solution onto a slowly rotating substrate in the spin coater while moving |

| | | | speed of pipette | over it with the pipetting channel |
|---|---|---|---|---|
| 16 | Stop spin coater | Spin coater | deceleration | Stop movement of spin coater |
| 17 | Move to take out position | Spin coater | angle | The spin coater rotates to the angle where the substrate can be picked up by the robot arm |
| 18 | Cap vial | Pipetting Robot | target vial | Close vial |
| 19 | Trash tip | Pipetting Robot | - | Put pipetting tip into trash bin |
| 20 | Pick up fresh cotton swab | Pipetting Robot | - | The pipette channel picks up a fresh cotton swab adaptor |
| 21 | Wipe edges | Pipetting Robot | - | Movement of the attached cotton swab over the substrate to wipe off stripes of the coated film |
| 22 | Put cotton swab back | Pipetting Robot | - | The cotton swab adaptor is put back into a rack of used adaptors, the cotton swabs in the adaptor will be replaced manually. |
| 23 | Transport substrate carrier to free waiting position | SCARA transport robot | carrier ID | The transport robot picks up the carrier by its ID and puts it to a free waiting position |
| 24 | Thermal annealing | Hot plate & SCARA transport robot | target Temperature, duration | The hot plate starts heating up to the target temperature. The transport robot then puts the substrate to the plate and leaves it there for the annealing duration |
| 25 | Cool down samples | SCARA transport robot | - | The hot substrates are transferred to a cold metal surface. |
| 26 | Take quality control image | Inspection camera & Light table | - | A picture of the whole substrate carrier is taken |
| 27 | Crop individual samples | *AMANDA* calculations | sample ID | The edges of the samples are detected and the samples are cropped and saved into the DB referenced to the sample IDs |
| 28 | Calculate layer homogeneity | *AMANDA* calculations | measurement ID | The texture of the cropped images of the samples is analyzed with the Gray-Level Co-Occurrence Matrix Method (GLCM) |
| 29 | Measure Absorbance | UV-VIS Spectrometer | sample ID, start wavelength, end wavelength, step width | Measures an absorption spectrum at certain points on the sample with the given parameters |
| 30 | Transfer carrier from Box2 to Box3 | SCARA transport robots | carrier ID | The carrier is moved through a sluice from Box2 to Box3 by both of the SCARA transpor robots |
| 31 | Put shadow mask to carrier | SCARA transport robots | carrier ID, mask ID | A shadow mask is put onto the substrate carrier by the transport robot |
| 32 | Turn carrier/mask stack | 6-axis robot | - | The stack of carrier/mask is turned around 180° |
| 33 | Load carrier/mask stack to evaporation holder | SCARA transport robots | - | The stack of carrier/mask is loaded to a holder for the evaporator |

| 34 | Open evaporator door | Evaporator | - | The flange door to the evaporator chamber is opened |
|---|---|---|---|---|
| 35 | Load evaporation holder into evaporator | 6-axis robot | - | The evaporation holder is loaded into the evaporator |
| 36 | Close Evaporator door | Evaporator | - | The flange door to the evaporator chamber is closed |
| 37 | Evacuate evaporation chamber | Evaporator | target pressure | The pumping procedure starts and evacuates the evaporation chamber. |
| 38 | Evaporate material | Evaporator | material, target thickness, evaporation rate, evaporation profile | The material is evaporated automatically. The rate is controlled by an internal PID control until the target thickness is reached |
| 39 | Vent evaporation chamber | Evaporator | - | Evaporation chamber is vented with nitrogen |
| 40 | Unload evaporation Holder from evaporator | 6-axis robot | - | The evaporation holder is unloaded from the evaporator |
| 41 | Unload carrier/mask stack to evaporation holder | SCARA transport robot | - | The stack of carrier/mask is taken out of the evaporation holder |
| 42 | take shadow mask from carrier | SCARA transport robot | - | The mask is taken away from the carrier by the SCARA transport robot |
| 43 | Turn on Solar simulator | Solar simulator | light intensity | The lamp of the solar simulator is turned on with target intensity |
| 44 | Drive sample into light | Linear stage | - | The substrates are located on a x-y-table |
| | beam | | | which moves the active sample into the middle of the light beam |
| 45 | Contact to sample | Linear stage | - | Electrical contact is made to the solar cells by a pin board |
| 46 | Measure current-voltage characteristics | SMU | start voltage, end voltage, step width | A current-voltage sweep of the given parameters is applied to the solar cell |
| 47 | Light induced degradation | LED light controller | light intensity, time, temperature | The substrates are put to a free degradation setup, light is turned on and the sample is degraded for the desired duration. |
| 48 | Calculate solar cell performance | *AMANDA* Calculations | solar cell area | $PCE$, $FF$, $J_{sc}$, $V_{oc}$, $R_{shunt}$ and $R_{series}$ are calculated from the IV-curve |
| 49 | Store substrate Carrier | SCARA transport robot | carrier ID | The carrier is transported to a storage tower |
| 50 | Deregister substrates and carrier | *AMANDA* control | carrier ID | The substrates and the carrier are set to a non-active state and are virtually removed from the facility. Substrate IDs are persistend, the measured data will be still available |

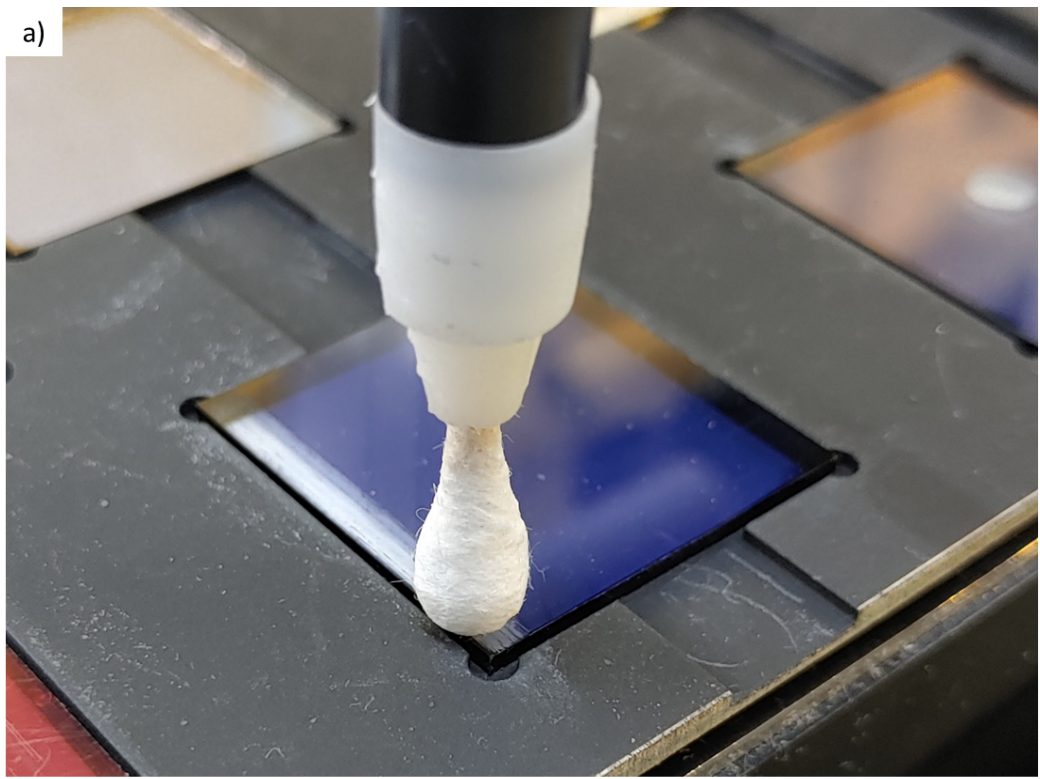

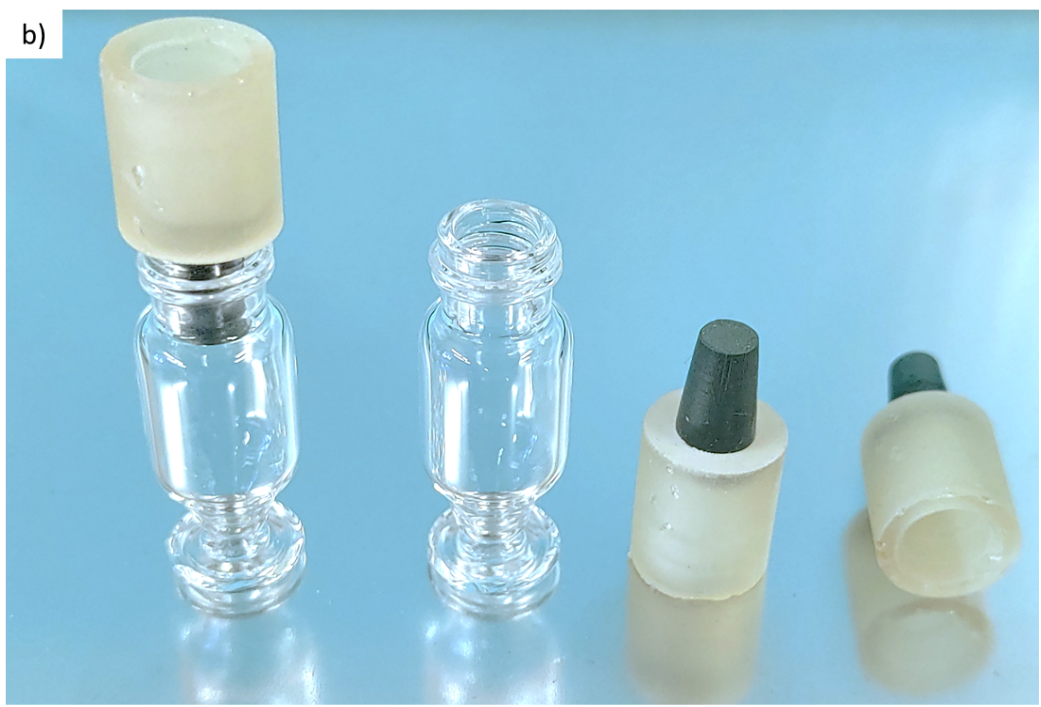

*Fig. S2: a) Cotton swap adaptor mounted onto the pipetting channel, removing a small stripe of the coated layer to contact the transparent back electrode. b) Left: 1 ml vial capped with a fluorinated rubber cork adaptor, which can be picked up by a pipetting channel of the pipetting robot. Middle and right: Bare vial and two cork adaptors.*

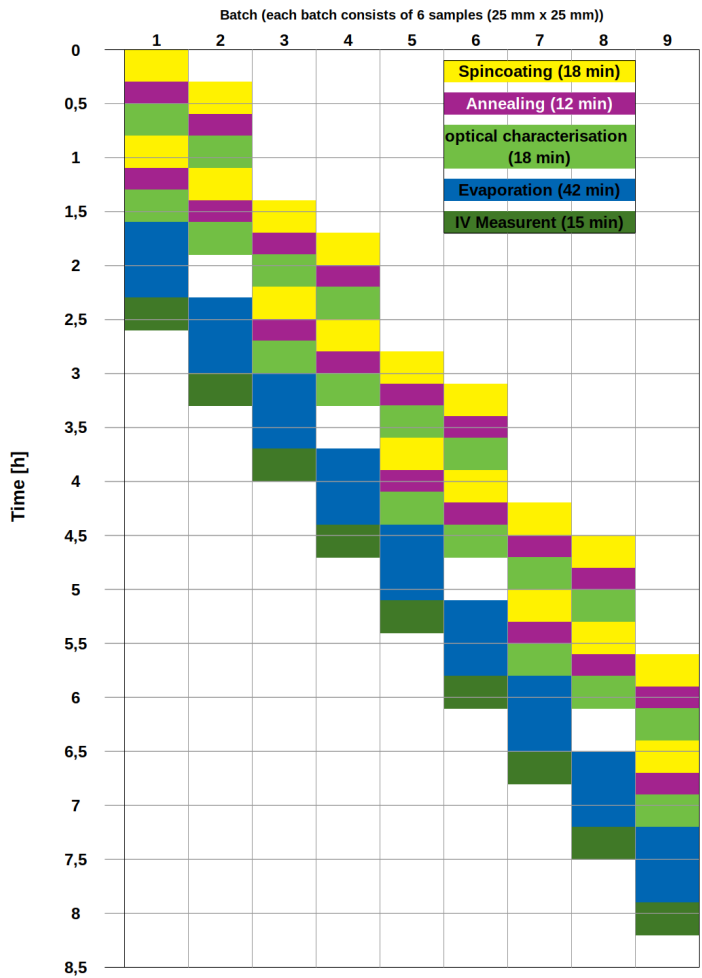

*Fig. S3: Parallel processing of OPV devices on 9 substrate carriers (batches) with either 6 or 8 substrates per carrier and 6 solar cells on each substrate.*

Figure S3 shows a theoretical throughput of sample batches on *AMANDA L1*. The processing time is shown on the y-axis, while the different batches are lined up beside each other on the x-axis. One batch consists of either 6 or 8 substrates which are carried throughout the machine on self developed substrate carriers. Each substrate holds 6 OPV devices. The processing of a single batch takes a bit over 2.5 hours, which would lead to a total of 3 processed batches in a normal working period of 8 hours. *AMANDA* however offers a parallelization system which allows *L1* to process different batches at different stations at the same time. The different stations are: spin coating, annealing, optical characterisation, evaporation and IV-measurement. With the help of this parallelization *L1* is able to reach cycle times of 42 minutes (time span of longest single task) for one batch. This results in a possible amount of 9 parallel processed batches in an 8 hour period, or even 34 batches in 24 hours assuming continuous production.